\documentclass[letter,
               keeplastbox,   % flushend option: not to un-indent last line in References
               ]{jacow}
\usepackage{pdfpages,multirow,ragged2e} %
\makeatletter%
	\ifboolexpr{bool{xetex}}
	 {\renewcommand{\Gin@extensions}{.pdf,%
	                    .png,.jpg,.bmp,.pict,.tif,.psd,.mac,.sga,.tga,.gif,%
	                    .eps,.ps,%
	                    }}{}
\makeatother

\ifboolexpr{bool{xetex} or bool{luatex}} % test for XeTeX/LuaTeX
 {}                                      % input encoding is utf8 by default
 {\usepackage[utf8]{inputenc}}           % switch to utf8

\usepackage[USenglish]{babel}

\ifboolexpr{bool{jacowbiblatex}}%
 {%
  \addbibresource{jacow-test.bib}
  \addbibresource{biblatex-examples.bib}
 }{}
\begin{document}

\title{Bayesian optimization scheme for the design of a nanofibrous high power target}

\author{W. Asztalos\thanks{wasztalos@hawk.iit.edu}, Y. Torun, Illinois Institute of Technology \\
S. Bidhar, F. Pellemoine, Fermi National Accelerator Laboratory \\
		P. Rath, Indian Institute of Technology Bhubaneswar
		%P. Contributor\textsuperscript{1}, Name of Institute or Affiliation, City, Country \\
		%\textsuperscript{1}also at Name of Secondary Institute or Affiliation, City, Country
		}

\maketitle

\begin{abstract}
   High Power Targetry (HPT) R\&D is critical in the context of increasing beam intensity and energy for next generation accelerators. Many target concepts and novel materials are being developed and tested for their ability to withstand extreme beam environments; the HPT R\&D Group at Fermilab is developing an electrospun nanofiber material for this purpose.
The performance of these nanofiber targets is sensitive to their construction parameters, such as the packing density of the fibers. Lowering the density improves the survival of the target, but reduces the secondary particle yield. Optimizing the lifetime and production efficiency of the target poses an interesting design problem, and in this paper we study the applicability of Bayesian optimization to its solution.
We first describe how to encode the nanofiber target design problem as the optimization of an objective function, and how to evaluate that function with computer simulations. We then explain the optimization loop setup. Thereafter, we present the optimal design parameters suggested by the algorithm, and close with discussions of limitations and future refinements.
\end{abstract}

\section{INTRODUCTION}
Neutrino beamlines are very useful for studying rare processes and Beyond the Standard Model physics. Many of these facilities operate by directing a primary beam of high energy protons at a fixed target; the higher the intensity of this primary beam, the higher the neutrino yield, which makes for more experimental data. Currently, maximum primary beam powers sit near the 1 megawatt threshold for the NuMI target at Fermilab, with future installations like the Long Baseline Neutrino Facility planned to exceed it in order to meet the demands for higher intensities.

The difficulty with realizing such a goal lies in the design of robust targets that can withstand the extreme environment created by a high intensity pulsed beam. Energy deposition from the beam leads to a rapid temperature rise which, combined with the pulsed nature of the beam, creates thermal stress. On top of that, the proton irradiation leads to defect formation, swelling, and embrittlement that further reduce the target's operational lifespan. Developing robust targets that can endure these challenges is the focus of the research area known as High Power Targetry (HPT), and many novel designs have been proposed. 

The HPT Research and Development (HPT R\&D) Group at Fermilab have been developing an electrospun nanofiber target concept which may be well-suited to the task \cite{productionQualification,nanofiberPoster}. These nanofiber ``mats'' are non-woven, and so the constituent fibers can move past one another, which could mitigate the damage from thermal stresses. The fibers also demonstrate resistance to radiation damage \cite{productionQualification}, which is thought to stem from the small grain size of the nanofiber crystal structure, as grain boundaries might be serving as good defect sinks. Cooling gas can also be forced through the porous mats, which may improve the cooling efficiency.

In 2018, the HPT R\&D Group organized an experiment at the HiRadMat Facility at CERN \cite{hrm1,hrm2} to perform thermal shock testing on target and beam window materials, with two samples of nanofiber targets included as part of their prototyping \cite{productionQualification}. Both samples were made of Yttria Stabilized Zirconia (YSZ), but differed in the packing density of the fibers, which we quantify via the \textit{Solid Volume Fraction} (SVF) of the target, denoted $f$. The SVF is the fraction of the target volume occupied by \textit{solid material}. Measurements of the samples in \cite{productionQualification} imply that the SVFs of the sample were ${f\approx 0.05}$ (low density sample) and ${f\approx 0.20}$ (high density sample).

\begin{figure}[!htb]
   \centering
   \includegraphics*[width=.9\columnwidth]{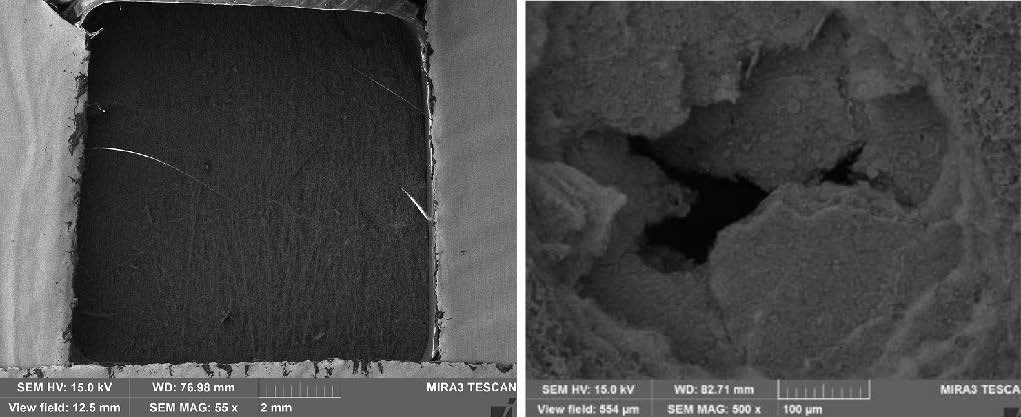}
   \caption{SEM images from HiRadMat test post-irradiation. Left: $f=0.05$ Right: $f=0.20$. Courtesy of Sujit Bidhar.}
   \label{fig:hrm}
\end{figure}

Following exposure to the high intensity pulsed proton beam at HiRadMat, examination of the samples revealed that the $f=0.05$ sample had no apparent damage, whereas the $f=0.20$ sample had a hole at the beam center, as seen in Fig~\ref{fig:hrm}. One of our hypotheses for this difference in performance is that the high density sample had a higher viscous resistance to fluid flow, and so when the air inside the mat was heated by the beam, it could not escape and exerted a greater pressure on the solid fibers, resulting in the sample blowing apart. This hypothesis is supported by our recent simulation work documented in \cite{Asztalos:IPAC24-TUPS44}.

This testing demonstrates that the performance of the nanofiber targets depends on their construction parameters. For nanofiber mats with large SVFs, backpressure formation threatens the survival of the target; however, lowering the SVF reduces the density of the target and hence the secondary particle yield drops. The loss can be recouped by using higher Z materials or increasing the target length, but only up to a point. As such, maximizing the survivability and yield of the nanofiber targets poses an interesting optimization problem. In this paper, we study whether classical Bayesian Optimization is a suitable tool for target design.

\section{Theory And Model}
\textit{Bayesian optimization} is a global optimization method of an objective function $\mathcal{F}$. The advantage of Bayesian Optimization (BO) is that very little information is required about $\mathcal{F}$. The principle behind BO \cite{bayesopt1} is that $\mathcal{F}$ is treated as a random function, described by a \textit{prior} probability distribution. Bayes' theorem is then applied to condition the prior distribution of $\mathcal{F}$ by the knowledge of its values at a finite collection of points---the resulting \textit{posterior} distribution is used to compute an \textit{acquisition function}, which is separately optimized in order to select the next point at which to evaluate $\mathcal{F}$. This process is continued to the user's tolerances for convergence. Certain choices of statistical models for priors and acquisition functions can ensure \textit{exponential} convergence to the global maximum \cite{bayesoptexp}.

We decided to use BO to help us design nanofiber targets; to do that, we needed to cast the trade-off between the potentially damaging pressure rise in the target and the secondary particle yield as the optimization of some objective function. The two (continuous) design parameters we can control at the moment are the SVF of the mat, $f$, and the average nanofiber radius, $R$, and so we use these as the input variables for $\mathcal{F}$, with bounds $f \in [0.05,\ 0.35]$ and $R \in [\SI{100}{nm},\ \SI{2.5}{\mu m}]$. These bounds are estimated from the limits of our manufacturing process and the expected threshold for a nanofiber target to become susceptible to thermal stress again. We would like to minimize the maximum pressure rise in the target, $\Delta P(f,R)$, while maximizing the secondary particle yield, $\mathcal{Y}(f)$, and so we elected to build $\mathcal{F}$ out of both a pressure and yield component, where each component is the relative difference from a reference value. As such, we choose for $\mathcal{F}$ to take the form:
\[ \mathcal{F}(f,R) \; := \; \left( \frac{\mathcal{Y}(f) - \mathcal{Y}_\text{ref}}{\mathcal{Y}_\text{ref}} \right) \; + \; \left( - \frac{\Delta P(f,R) - P_\text{ref}}{P_\text{ref}} \right)\]
where ${\mathcal{Y}_\text{ref}:=\SI{2.89}{}\times 10^{-3}}$~(non-proton charged hadrons Per Primary Proton (PPP)) is chosen to be the calculated secondary particle yield of an equivalent target made of graphite, the conventional target material in neutrino beamlines, so that lowering yield from the graphite equivalent is ``penalized''. We chose the reference pressure $P_\text{ref}=\SI{100}{kPa}$ by rounding up the pressure rise calculated for air inside the low density sample at HiRadMat \cite{Asztalos:IPAC24-TUPS44}, $\SI{67}{kPa}$, since that target survived and thus exceeding its pressure rise should be ``penalized''. This choice of $\mathcal{F}$ is delicate, since it is the connection between the mathematics and the physical problem, but this was decided as a natural choice since fractional depreciation in one criterion oppose the improvement of the other directly. Optimizing $\mathcal{F}$ thus offers one way to approach the trade-off in designing nanofiber targets.

\section{METHOD}
To evalute $\mathcal{F}(f,R)$, we first use MARS \cite{MARS1,MARS2,MARS3} to calculate the secondary particle yield and energy deposition by the HiRadMat beam into a \SI{10}{mm} $\times$ \SI{10}{mm} $\times$ \SI{0.1}{mm} YSZ nanofiber mat with construction parameters $(f,R)$. The beam parameters from the 2018 testing were \SI{440}{GeV} protons with RMS size $\sigma=\SI{0.25}{mm}$, a pulse length of $\SI{4}{\mu s}$, and $1.21 \times 10^{13}$ protons on target. The yield $\mathcal{Y}(f)$ is estimated by counting the number of charged hadrons through the back of the target PPP and subtracting $\exp(-L/I(f))$, where $L$ is the thickness of the target ($\SI{0.1}{mm}$) and $I(f)$ is the estimated nuclear interaction length of the target (taken to be the mass fraction weighted average of the interaction lengths of the constituent materials \cite{pdg}). This is to remove non-interacting protons from the charged hadron count, giving a secondary particle count PPP. The energy deposition data is fit to a Gaussian in the radial distance from the target center and incorporated as a volumetric heat source for the calculation of $\Delta P(f,R)$.

\begin{figure}[!htb]
   \centering
   \includegraphics*[width=.9\columnwidth]{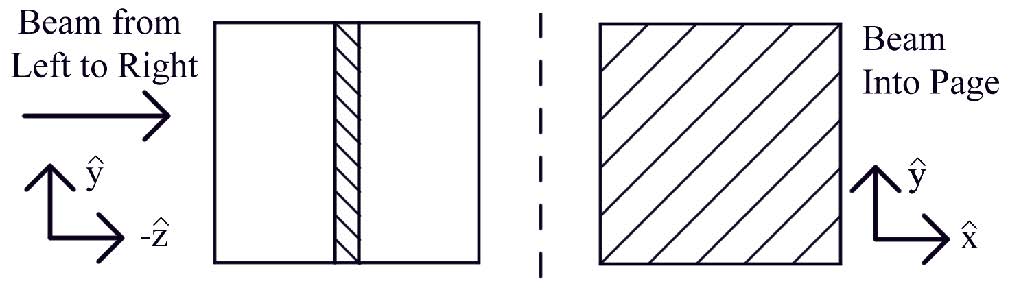}
   \caption{Illustration of problem domain (not to scale). Hatching indicates the nanofiber porous zone.}
   \label{fig:domain}
\end{figure}

The maximum pressure rise, $\Delta P(f,R)$, is obtained from a multiphysics simulation using ANSYS Fluent \cite{fluent}, taken to be the maximum air pressure in the target center computed during the $\SI{40}{\mu s}$ simulation time. To handle the nanostructure of the target, effective thermal conductivities are assigned \cite{insulation,Bhattacharyya} for the thermal model and Darcy's law is used to handle the air flow, with permeability to fluid flow, $\alpha(f,R)$, taken from \cite{permeability}. The solution domain is illustrated in Fig.~\ref{fig:domain}, and the Boundary Conditions (BC's) for the energy equation are zero heat flux for the $\pm \hat{x}$ and $\pm \hat{y}$ walls and fixed temperatures of $\SI{300}{K}$ at the $\pm \hat{z}$ boundaries. The momentum equation BC's are zero velocity at the $\pm \hat{x}$ and $\pm \hat{y}$ walls, and zero gauge pressure at the $\pm \hat{z}$ outlets. The energy and momentum equations are solved with the SIMPLE algorithm, with a timestep of ${\Delta t=\SI{0.1}{\mu s}}$ during the $\SI{4}{\mu s}$ beam pulse and ${\Delta t=\SI{0.5}{\mu s}}$ afterward. The full simulation details can be found in \cite{Asztalos:IPAC24-TUPS44}, and more explanation of the theory in \cite{ms}.

This constructs $\mathcal{F}(f,R)$. The actual Bayesian optimization loop was implemented using the BoTorch python library \cite{botorch}. We chose Gaussian processes as our priors for $\mathcal{F}$, and minimized the negative marginal log likelihood to train the model hyperparameters. We used Expected Improvement as our acquisition function, since it features an automatic exploration and exploitation trade-off, and we did not have time to trial trade-off parameters with an Upper Confidence Bound acquisition function for this contribution.

\section{RESULTS}
\begin{figure}[!htb]
   \centering
   \includegraphics*[width=.9\columnwidth]{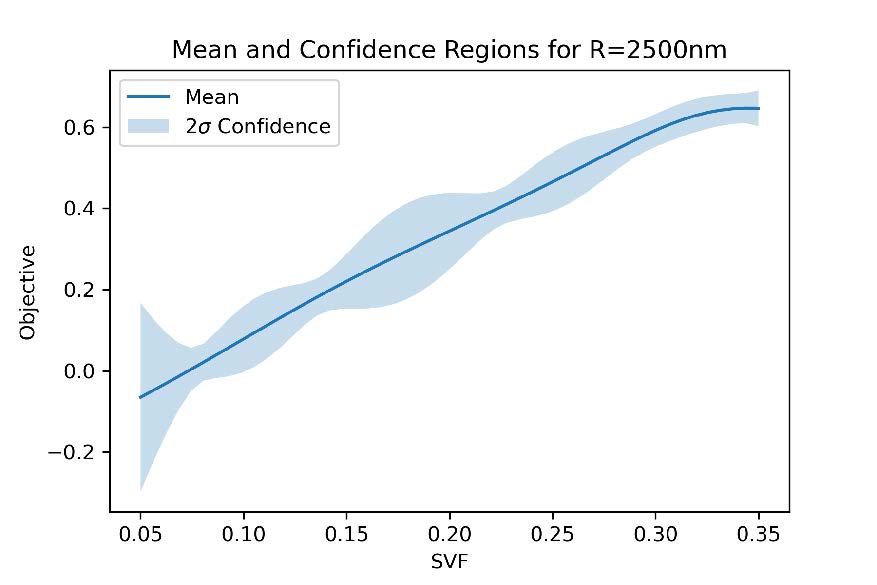}
   \caption{Final iteration mean value and $2\sigma$ confidence intervals of posterior distribution of $\mathcal{F}$ for $R=\SI{2500}{nm}$.}
   \label{fig:meanconfidence}
\end{figure}

The optimizer was trained on three initial data points---the ${f=0.05}$ and ${f=0.20}$, ${R=\SI{156.5}{nm}}$ results that emulate the HiRadMat test \cite{Asztalos:IPAC24-TUPS44}---and ${f=0.10}$ and ${R=\SI{1000}{nm}}$. The optimizer was then run for 13 iterations until convergence behavior was demonstrated: ${R=\SI{2500}{nm}}$ was quickly identified as optimal. Figure~\ref{fig:meanconfidence} depicts the mean and confidences of the posterior distribution on $\mathcal{F}$ as a function of $f$ with $R$ fixed at its optimal value of $\SI{2500}{nm}$. We see that the space is suitably explored to identify ${f=0.35}$ as the optimal SVF.

\begin{figure}[!htb]
   \centering
   \includegraphics*[width=.9\columnwidth]{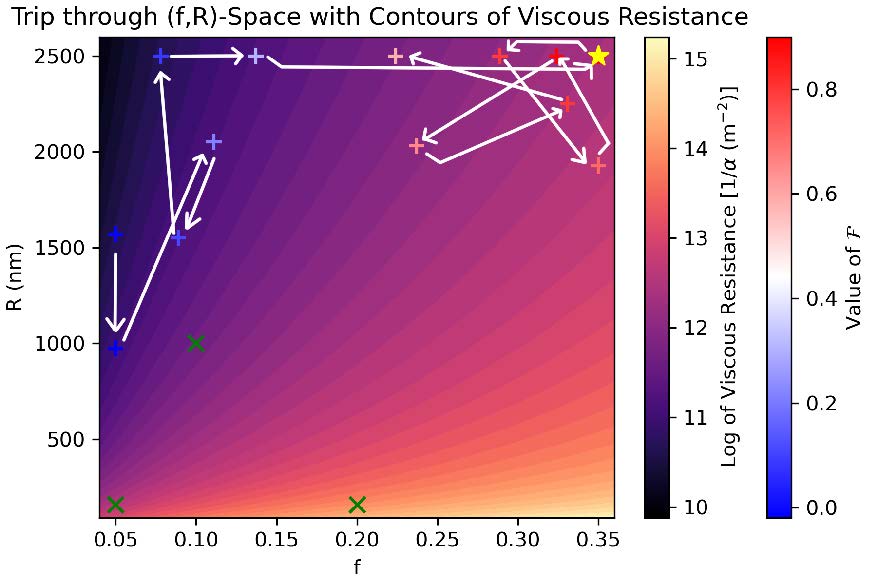}
   \caption{Illustration of optimizer's trip through parameter space. Background is contours of common log of $\alpha(f,R)$.}
   \label{fig:paramspace}
\end{figure}

Figure~\ref{fig:paramspace} depicts the trip taken through $(f,R)$ parameter space. Green x's are the training points, +'s are the points selected by the optimizer which are colored according to the value of $\mathcal{F}$, and the background is contours of the logarithm of the viscous resistance, ${1/\alpha}$. The best point is indicated by a gold star. We see that the optimizer initially favored small values of $f$ to lower $\Delta P$, but eventually ``discovered'' that increasing $R$ improves the permeability and thus lowers $\Delta P(f,R)$ without changing $f$. It then tested out increasing $f$ to improve the yield $\mathcal{Y}(f)$, but kept $R$ large to avoid increasing $\Delta P$.

% Figure~\ref{fig:criterionplot} depicts the value of the objective function, $\mathcal{F}$, as well as the value of the separate pressure and yield components as a function of the iteration.

% \begin{figure}[!htb]
%    \centering
%    \includegraphics*[width=.9\columnwidth]{TUPS45/BOCriterionPlots.png}
%    \caption{Plot of $\mathcal{F}$, along with the pressure and yield components.}
%    \label{fig:criterionplot}
% \end{figure}

\section{DISCUSSION}
The path of the optimizer through $(f,R)$ space shows that increasing $R$ for a fixed $f$ can lower $\Delta P$ and potentially improve target lifetime for the same yield. It converged to a value of ${R=\SI{2500}{nm}}$ and ${f=0.35}$. The fact that these are the maximum values we \textit{allowed} for the optimizer to use is troubling, since it implies that it would increase $R$ further if it could. Ultimately, this stems from a lack of penalization of large values of $R$ in the current objective function. The problem is that for large $R$ the energy deposition in individual fibers would increase and potentially reintroduce the thermal stresses that the nanofiber target was designed to avoid. Since $\mathcal{Y}(f)$ is independent of $R$, though, and $\alpha \sim R^2$ \cite{permeability}, with a larger $\alpha$ giving a smaller $\Delta P$ \cite{darcy}, the optimizer will favor the largest value of $R$. Slightly more interesting is that the max value of $f$ was settled on too, since increasing $f$ decreases $\alpha$. This does indicate that the ``trade-off'' was still tested by this objective function, but that the model is still missing other information about the effect of $R$, which makes the suggested optimal parameters doubtful.

\section{CONCLUSION}
The optimum that the Bayesian optimizer converged to, $f=0.35$, $R=\SI{2500}{nm}$ coincides with the maximum values we allowed for these variables. This implied runaway behavior is entirely because there was no penalization of large $R$, and in retrospect this stems from not including a thermal or mechanical objective, which could be devised. Another option for future work is to move from classical BO to multi-objective Bayesian optimization. In this setting, we can instead generate Pareto frontiers, and decide upon values of $f$ and $R$ along a continuum of optimal values, which provides more options and utility from a design standpoint. Nonetheless, we have seen that the BO procedure, which chooses points to test the performance of a target at conservatively, works well with this kind of problem class. With room for refinements, and some promise already demonstrated, we are confident that this can be honed into a useful tool for designing high power targets.

\section{ACKNOWLEDGMENTS}
This work was produced by Fermi Research Alliance, LLC under Contract No. DE-AC02-07CH11359 with the U.S. Department of Energy, Office of Science, Office of High Energy Physics. The research presented here was possible with the support of the Fermilab Accelerator PhD Program.

%
% only for "biblatex"
%
\ifboolexpr{bool{jacowbiblatex}}%
	{\printbibliography}%
	{%
	% "biblatex" is not used, go the "manual" way
	
	%\begin{thebibliography}{99}   % Use for  10-99  references

    % \item[{[1]}] S. Bidhar \textit{et al.}, ``Production and qualification of an electrospun ceramic nanofiber material as a candidate future high power target'', \textit{Phys. Rev. Accel Beams}, vol. 24, p. 123001, Dec 2021. \vspace{-0.5em}
    % \item[{[2]}] S. Bidhar, ``Electrospun nanofiber materials for high power target applications'', tech. rep., Fermi National Accelerator Lab, 2017.\vspace{-0.5em}
    % \item[{[3]}] K. Daryabeigi, ``Heat transfer in high-temperature fibrous insulation'', \textit{Journal of Thermophysics and Heat Transfer}, vol. 17, no. 1, pp. 10-20, 2003.\vspace{-0.5em}
    % \item[{[4]}] R. Bhattacharyya, ``Heat-transfer model for fibrous insulations'', \textit{ASTM special technical publications}, pp. 272-286, 1980.\vspace{-0.5em}
    % \item[{[5]}] M. Tomadakis and T. Robertson, ``Viscous permeability of random fiber structures: Comparison of electrical and diffusional estimates with experimental and analytical results'', \textit{Journal of Composite Materials}, vol. 39, no. 2, pp. 163-188, 2005.\vspace{-0.5em}
    % \item[{[6]}] N.V. Mokhov and C.C. James, \textit{The MARS code system user's guide, version 15 (2016)}. Fermilab-FN-1058-APC, 2017.\vspace{-0.5em}
    % \item[{[7]}] N. Mokhov \textit{et al.}, ``MARS15 code developments driven by the intensity frontier needs'', \textit{Prog. Nucl. Sci. Technol.}, pp. 496-501, 2014.\vspace{-0.5em}
    % \item[{[8]}] N. Mokhov, ``Status of MARS code'', Fermilab-Conf-03/053, (Batavia, Illinois), 2003.\vspace{-0.5em}

} % end \ifboolexpr

%
% for use as JACoW template the inclusion of the ANNEX parts have been commented out
% to generate the complete documentation please remove the "%" of the next two commands
% 
%%%%\newpage

%%%%\include{annexes-Letter}

\end{document}